\documentstyle[psfig,twocolumn,aps,graphicx,rotating,epsfig,amsfonts]{revtex}
\tighten
\begin{document}

\setlength{\unitlength}{1cm}

\draft

\title{\bf Triplet Dispersion in CuGeO$_3$: Perturbative Analysis}

\author{Christian Knetter\thanks{
e-mail: ck@thp.uni-koeln.de
\hspace*{\fill} {\protect\linebreak}
internet: www.thp.uni-koeln.de/\~{}ck/} and G\"otz S. 
Uhrig\thanks{e-mail: gu@thp.uni-koeln.de
\hspace*{\fill} {\protect\linebreak}
internet: www.thp.uni-koeln.de/\~{}gu/}}

\address{Institut f\"ur Theoretische Physik, Universit\"at zu
  K\"oln, Z\"ulpicher Str. 77, D-50937 K\"oln, Germany\\[1mm]
  {\rm(\today)} }

\maketitle

\begin{abstract}
We reconsider the 2d model for CuGeO$_3$ introduced 
previously (Phys.~Rev.~Lett.~79, 163 (1997)). 
Using a computer aided perturbation method based
on flow equations we expand the 1-triplet dispersion up to
10$^{\rm th}$ order. The expansion is provided as a polynom in the
model parameters. The latter are fixed by fitting
the theoretical result to experimental data obtained by INS. For a
dimerization $\delta\approx 0.08(1)$ we find an excellent agreement with
experiment. This value is at least 2 to 3 times higher
than values deduced previously from 1d chain approaches. For the
intrachain frustration $\alpha_0$  we find
a  smaller value  of $0.25(3)$.
The existence of interchain frustration conjectured previously is
confirmed by the analysis of temperature dependent susceptibility.
\end{abstract}


\narrowtext

\section{Introduction}
The dispersion of the magnetic excitations is
an important source of information on experimental low-dimensional
spin systems. Knowledge of the dispersion relation
$\omega(\vec{k})$  helps essentially to identify 
the model appropriate to describe the compound under study.
The dispersion relation provides also important insight in
the nature of the ground state. Very common in low dimensional
systems is the scenario
of a singlet $S=0$ ground state {\em without} magnetic long range order
(a ``spin liquid'') of which the elementary excitations
are triplets $S=1$. These systems are generically gapped.
Examples are isolated or weakly coupled 
dimerized spin chains and spin ladders such as 
(VO)$_2$P$_2$O$_7$,\cite{garre97a} the spin-Peierls phase of 
CuGeO$_3$,\cite{bouch96,uhrig97a}
 and SrCu$_2$O$_3$. \cite{azuma94}
A true 2D example is SrCu$_2$(BO)$_2$ which is characterized
by frustrated dimers.\cite{kagey99,miyah99,mulle00a}

In these gapped $S=1/2$ systems where the gap is related to 
some ``strong'' bond (which can be also the rung of a 2-leg ladder)
the elementary triplet excitations are in principle accessible
by a perturbative expansion about the  limit of isolated dimers.
This approach, however, becomes tedious for the description of
realistic materials since  the expansion parameter is often not
really small. Thus one has to compute high orders to achieve
quantitative agreement. For this reason various automated
approaches have been conceived which leave the tedious part to
computers \cite{he90,gelfa90,barne99,knett00a}.

\begin{figure}
\includegraphics[width=8cm]{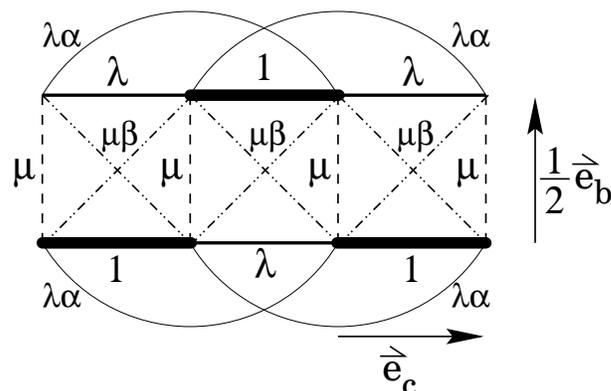}
\caption{
Dimerization pattern in the dimerized low temperature phase of CuGeO$_3$.
The couplings are denoted relative to the strongest coupling $J$ which
is set to unity in the figure.
\label{koppl1}}
\end{figure}

In the present article, we will apply the previously introduced
perturbation by flow equation \cite{knett00a} to the two-dimensional,
though anisotropic, system of CuGeO$_3$ in its dimerized low-temperature
phase \cite{bouch96}. Thereby we extend the previous
analysis (Ref.~\cite{uhrig97a}, henceforth cited as [I]) considerably. 
Our starting point remains the same as before. The strongest
coupling is given by $J$; the other couplings are given relative
to $J$ as indicated in Fig.~\ref{koppl1} (for details see
Fig.~1 in [I]).

\section{Method}
The problem to be solved reads
\begin{equation}
H = H_0 + \lambda H_S\ .
\end{equation}
As in the chain in Ref.~\cite{knett00a} the isolated dimer 
limit ($\lambda =0$ at
finite $\mu/\lambda$) has an equidistant
energy spectrum and the perturbation can alter the number of energy
quanta (here: triplets on the dimers) 
by 2 at maximum. Hence $H_S$ can be represented
as $H_S=T_{-2}+T_{-1}+T_0+T_1+T_2$ where $T_i$ stands for the perturbing part
changing the number of elementary triplets by $i$.
The same formalism as in Ref.~\cite{knett00a} can be used. 
This formalism maps the perturbed Hamiltonian by a
continuous unitary transformation, the so-called flow equation method
\cite{wegne94}, to an effective Hamiltonian $H_{\rm eff}$ which
{\em conserves} the number of energy quanta, i.e. $0=[H_{\rm eff},H_0]$.
The effective Hamiltonian has the form
\begin{equation}
\label{H_eff}
H_{\rm eff} = H_0 +\sum_{k=1}^{\infty}\lambda^{k} 
\sum_{|\underline{m}|=k, M(\underline{m})=0} C(\underline{m}) 
T(\underline{m})\ ,
\end{equation}
where $\underline{m}$ is a vector of dimension $k$ of which the
components are in $\{\pm 2,\pm 1,0\}$; $M(\underline{m})=0$ signifies
that the sum of the components vanishes which reflects the conservation of the
number of energy quanta (triplets). The coefficients 
$C(\underline{m})$ are generally valid fractions computed in 
Ref.~\cite{knett00a} where also further
details on the flow equation method can be found.

\begin{figure}[hb]
\includegraphics[width=8.8cm]{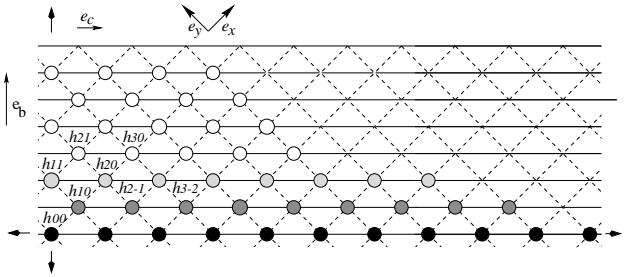}
\caption{Effective lattice on which the triplet hops. 
We calculate the amplitudes $h_{j,l}$ for all hopping processes
starting on $(0,0)$ and ending on one of the depicted dimers
(circles). For all circles that are accessible by an arbitrary hopping 
of length six or less the amplitudes have been calculated within 6$^{\rm th}$
 order. (The length of a hop is the minimum number of bonds (solid or dashed)
required to link start and end point.)
The
amplitudes for light gray, dark gray and black circles have been
extended in 8$^{\rm th}$ order, provided that these sites
can be reached by a hopping $\propto \mu^2$ of length 8.
Analogously, the amplitudes for dark gray and black circles
were extended by calculating hopping processes $\propto \mu^1$
within 9th order. Finally, the amplitudes for all black circles were
extended by processes $\propto \mu^0$ within 10th order. The arrows
indicate axes with respect to which reflection symmetry holds.}
\label{eff-latt}
\end{figure}

Since $H_{\rm eff}$ conserves the triplet number the one-triplet
sector is particularly easy to solve. Acting on one triplet the action of 
$H_{\rm eff}$ may only consist in shifting the triplet. This means
that the triplet hops on 
an effective lattice where one site stands for one dimer on the
original lattice, see Fig.~2 in [I] or Fig.~\ref{eff-latt}.
The full dispersion $\omega(\vec{k})$ is obtained by Fourier transform
\begin{equation}
\label{disp}
\omega(\vec{k}) = J\sum_{j,n} h_{j,n}\exp(i(k_1 j+k_2 n))\ .
\end{equation}

\begin{figure}[hb]
\begin{picture}(5.2,5)\put(1.5,0.2){
\epsfig{file=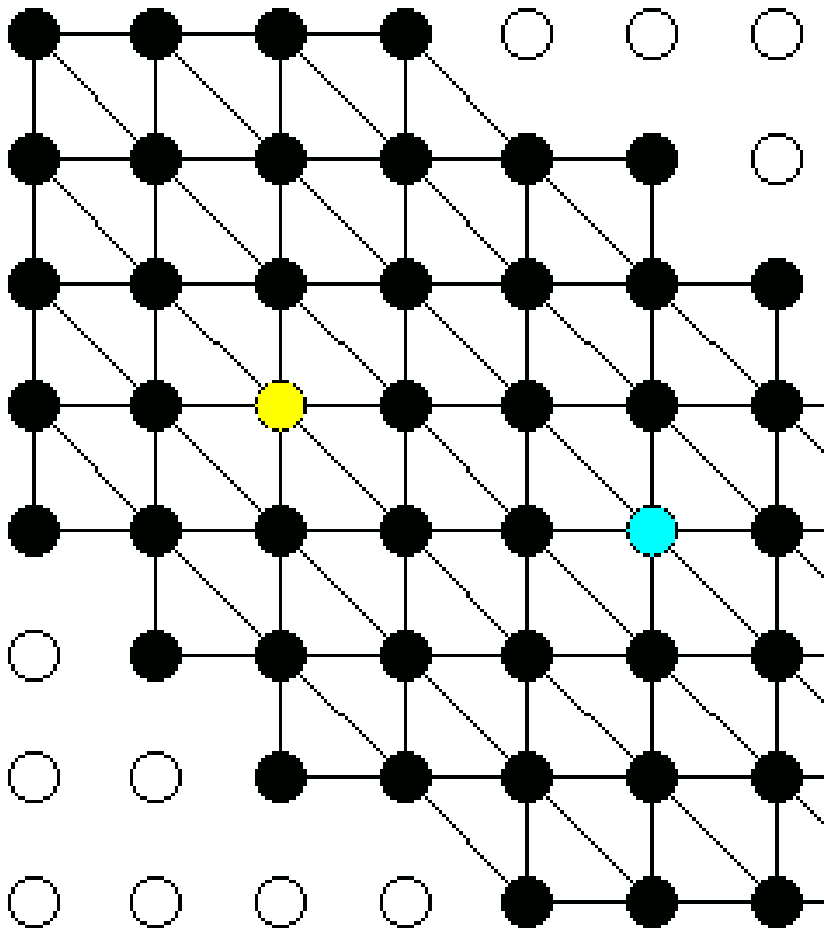,width=3.6cm,height=3.6cm,
angle=45,silent=,clip=}
}\end{picture}
\caption{Computer generated cluster necessary to compute $h_{3,-1}$ in
8$^{th}$ order, allowing for arbitrary hopping processes of length 8.
 The light gray (dark gray) circle denotes the start (end) 
dimer $(0,0)$ ($(3,-1)$).}
\label{cluster}
\end{figure}

The hopping amplitudes $h_{j,n}$ can be calculated on finite clusters of the
 (in principle infinite) effective lattice: From the linked cluster
theorem we know that the finite order contribution of a short-ranged
perturbation does not depend on the cluster size for sufficient large
clusters. Carrying out our perturbation within order $l$ implies
that one allows dimer to dimer hopping processes of length $l$
\cite{knett00a}. The minimum cluster for a given
amplitude $h_{j,n}$ in a given order $l$ contains all dimers and
links that are  involved in a  hopping of
length $\le l$ starting at dimer $(0,0)$ and ending at 
$(j,n)$. The minimum cluster is determined by considering all paths from
$(0,0)$ to $(j,n)$. All dimers and links covered by one of these paths
are part of the minimum cluster.
In Fig.~\ref{cluster}, the computer generated
minimum cluster for calculating $h_{3,-1}$ in order 8 is shown.

Due to the strong anisotropy of the quasi-1D system CuGeO$_3$
it is reasonable to use higher order terms only along the chains.
This simplifies the computational task considerably 
since the calculation of a hopping
process along the chain is much simpler. The cluster
to be considered can be chosen smaller. The same is true for
hopping processes {\em close} to the chain direction. Here we restrict
the hopping processes to be at maximum quadratic in the interchain hopping
$\mu$, which reduces the
cluster sizes significantly so that the perturbation order can be
enlarged.

\section{Analysis of Experimental Data}
The results for the $h_{j,n}$ are too lengthy to be published in
written form. We will provide them in electronic form on our
home pages on appearance of this article. In~[I] the
$h_{j,n}$ in third order in $\lambda$ and $\mu$ were presented.
A few of these are erroneous. They are corrected herewith \cite{note-corr}.
The corrections, however, have no influence on the conclusions in [I]
(see also discussion below).

Once all amplitudes $h_{j,n}$ are calculated the dispersion relation
is given by Eq.~(\ref{disp}). After rewriting
Eq.~(\ref{disp}) in terms of $k_b$ and $k_c$ (the reciprocal basis to
$e_b$ and $e_c$) we add the term $4t_a \cos(k_a)\cos(k_c)$ with 
$4t_a=0.22$ meV to account for the dispersion
in a-direction (cf.~[I]). To fix the parameters
$J$, $\alpha$, $\beta$, $\mu$ and $\lambda$ (cf.~Fig.~\ref{koppl1}) we use
the one-magnon dispersion data for CuGeO$_3$ experimentally determined
by inelastic neutron scattering \cite{regna96a}. Note
that  the hopping amplitudes are computed 
as polynomials over $\mathbb{Q}$ in the
parameters. 

\begin{figure}[hb]
\includegraphics[width=6cm]{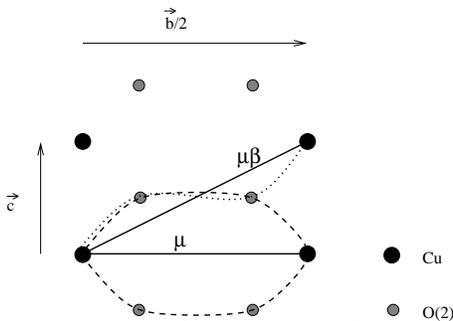}
\caption{Schematic view of the microscopic super-exchange paths between the 
chains (running along the c-direction) in CuGeO$_3$. The three-dimensional
situation is depicted in Ref.~\protect\cite{brade96a}.} 
\label{boris}
\end{figure}

As  noticed in [I] the parameter $\beta$ has almost
no influence on the shape of the dispersion. Hence we refrain from
determining $\beta$ from the dispersion but set it beforehand to 
some reasonable values in the interval $[-0.3,0.3]$. This
choice  is motivated by comparing the
microscopic {\em direct} super-exchange path $\mu$ and the 
{\em shifted} super-exchange
path $\mu\beta$ shown schematically in Fig.~\ref{boris} 
( cf.~Fig.~\ref{koppl1}). 
There is only one
path per Cu$^{2+}$-site for the shifted coupling whereas there are two paths
for the direct coupling. Thus we expect $|\mu\beta|\approx 1/2 |\mu|$,
i.e.\ $|\beta|\approx 0.5$. There are also results from ab-initio
calculation for the interchain couplings which indicate the existence
of interchain frustration \cite{rosne00}. Further evidence is provided below
by the analysis of the susceptibility.
Furthermore, we find that for $|\beta| > 0.4$ fits to the dispersion data
become worse.

\begin{figure}[hb]
\includegraphics[width=8.8cm]{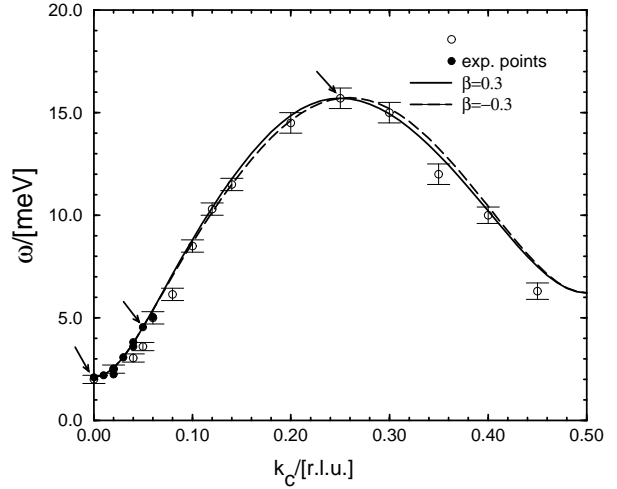}
\caption{Dispersion $\omega(k_a=0,k_b=0,1/2+k_c)$ in
  c$^*$-direction. The arrows indicate the experimental points used to fix
  the parameters. $10^{th}$ order fits based on $(\beta, \lambda,
  \alpha, \mu, J)$ = (0.3, 0.836, 0.225, 0.266, 13.1meV)
  and (-0.3, 0.846, 0.209, 0.081, 12.3meV), respectively.} 
\label{disp_big_c}
\end{figure}

\begin{figure}[hb]
\includegraphics[width=8.8cm]{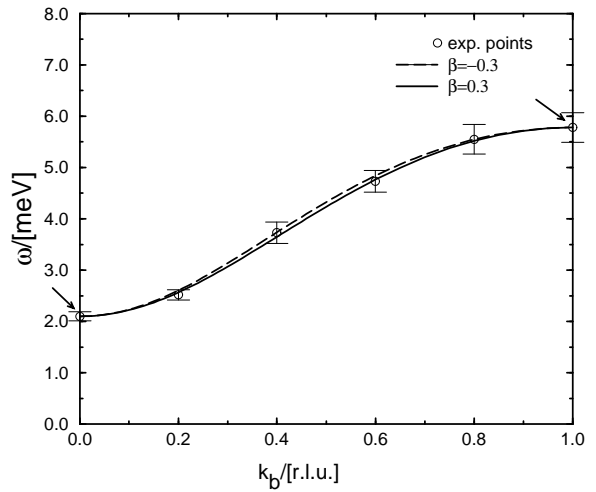}
\caption{Dispersion $\omega(k_a=0,k_b,k_c=0)$ in
  b$^*$-direction. Otherwise as in Fig.~\ref{disp_big_c}}
\label{disp_big_b}
\end{figure}

 To determine the remaining parameters we
equate four different experimental points with the corresponding
parameter dependent dispersion values given by Eq.~(\ref{disp}). The
parameters are fixed by solving the resulting system of
equations.  For $\beta = 0.3$ and $-0.3$
Figs.~\ref{disp_big_c},~\ref{disp_big_b} show the resulting  
dispersion curves in c$^*$- and in b$^*$-direction, respectively, using all
$h_{j,n}$ calculated.

\begin{figure}[hb]
\includegraphics[width=8.8cm]{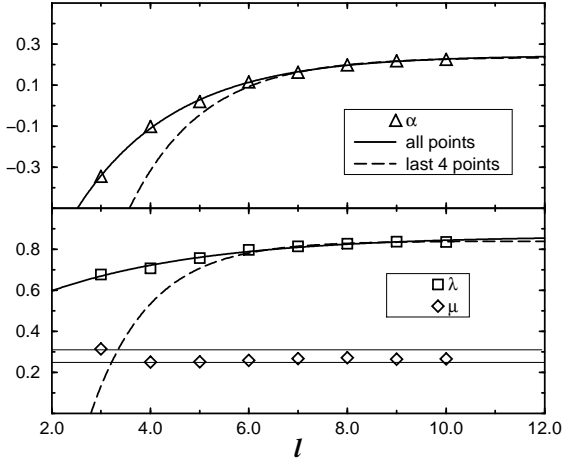}
\caption{Dependence of the parameter values on the
  perturbation order $l$ at $\beta=0.3$. The
  quasi-constant behavior of $\mu$ is found for all $\beta$-values
  checked. The lines are fits to the data according to
  Eq.~(\ref{extrapol}). The solid lines consider all points, the
  dashed ones only the last four points.} 
\label{trend_1}
\end{figure}

As can be seen from Figs.~\ref{disp_big_c} and \ref{disp_big_b}
the plain series up to $10^{\rm th}$ order provides excellent fits. Yet
one realizes that the parameter values still change on passing from
order to order. So it appears that even at $10^{\rm th}$ order the
results are not quantitative. In order to obtain quantitative reliable
results we adopt a systematic extrapolation in the
order. In each order $l \in\{3,4,\ldots10\}$ we determine the optimum
fit parameters. For illustration, Fig.~\ref{trend_1} shows results for
 $\beta=0.3$. 

Assuming exponential convergence we use
\begin{equation}
\label{extrapol}
f(l)=X-be^{-2cl}\ ,
\end{equation}
where $X$ is the asymptotic value of the parameter considered and $b$ and
$c$ are constants. The choice (\ref{extrapol}) is motivated on one hand
by its obvious applicability (see Fig.~\ref{trend_1}).
 On the other it stems from the fact that
 CuGeO$_3$ is a quasi one-dimensional gapped spin system. So one expects the
magnetic correlations to drop exponentially with distance. 
Furthermore the order $l$ determines the maximum distance over which
correlations occur (cf.~Ref~\cite{knett00a}, 
namely $l$ counted in dimers or $2l$ counted in spin sites.
 Hence the constant $c$ in Eq.~(\ref{extrapol}) can be
understood as the inverse of a correlation length $\xi$. 
With the usual relation $\xi \approx v_S/\Delta$ for one dimensional systems
we obtain $c\approx 1/6$ based on the rough estimates $v_S=\pi/2\cdot
J(1-1.12\alpha_0)$ \cite{fledd97} and $\alpha_0\approx 0.3;
 J \approx 12\mbox{meV}; \Delta \approx 2\mbox{meV}$.
This is indeed what is found (cf. Tab.~I,~II) so that we judge our 
extrapolations as being well justified.

In Fig.~\ref{trend_1} the extrapolations are depicted
by lines. The solid lines were obtained by using all calculated
parameter values. The dashed lines are obtained from on 
the last four points, i.e.\ the results in order 7, 8, 9, and 10. 
The deviation between these two 
extrapolations are used as a measure for the extrapolation error.
This procedure is carried out for $\alpha, \lambda$, and $J$.
 
There is no systematic dependence of $\mu$ on the order
$l$. The parameter $\mu$ oscillates between the two
thin horizontal lines in Fig.~\ref{trend_1}. So we take the
average of these two bounds as our estimate for $\mu$ and their
difference as the error in the determination of $\mu$.

Tabs.~I and~II summarize the results of the fits for the parameters
$\alpha$, $\lambda$ and $J$ for four preset values of $\beta$.
The values for $\mu$ are listed in  Tab.~III. 

\begin{center}
    \footnotesize{ 
      \begin{tabular}[t]{|c||c|c|c||c|c|c|}
      \hline
      &\multicolumn{3}{|c||}{All points
      considered}&\multicolumn{3}{|c|}{last four points
      considered}\\\hline 
      &$X$&$b$&$c$&$X$&$b$&$c$\\\hline\hline
      \multicolumn{7}{|c|}{$\beta=0.3$}\\\hline
      $\alpha$ &0.245&2.61&0.249&0.236&8.24&0.338\\\hline
      $\lambda$&0.867&0.501&0.155&0.840&8.66&0.418\\\hline
      $J/$meV   &13.6&11.0&0.164&13.2&105&0.357\\\hline\hline
      \multicolumn{7}{|c|}{$\beta=0.22$}\\\hline
      $\alpha$ &0.232&3.27&0.268 & 0.228&9.50&0.343\\\hline
      $\lambda$&0.863&0.681&0.184 & 0.842&13.6&0.444\\\hline
      $J/$meV   &13.1&11.3&0.171 & 12.8& 68.9 &0.324\\\hline\hline
      \multicolumn{7}{|c|}{$\beta=0$}\\\hline
      $\alpha$ &0.218&4.16&0.294&0.226&9.20&0.334\\\hline
      $\lambda$&0.862&0.902&0.213&0.848&14.94&0.442\\\hline
      $J/$meV   &12.8&11.1&0.174&12.6&44.48&0.287\\\hline\hline
      \multicolumn{7}{|c|}{$\beta=-0.3$}\\\hline
      $\alpha$ &0.212&4.48&0.300&0.222&9.94&0.337\\\hline
      $\lambda$&0.863&0.974&0.218&0.849&16.5&0.448\\\hline
      $J/$meV   &12.7&10.9&0.173&12.5&42.2&0.282\\\hline
    \end{tabular}}
\end{center}
\ TABLE I.\small{ Extrapolated parameter values $X$ according to
  Eq.~(\ref{extrapol}). The
  experimental points we used in the fit process for this table
  are (cf. Figs~\ref{disp_big_c},~\ref{disp_big_b})
  [$(k_b,k_c);\omega(\mathbf{k})/ \mbox{meV}$]:\newline
  [$(0,0); 2.1$], [$(0,0.05); 4.55$], [$(0,0.25); 15.7$], [$(1,0);5.78$].} 
\begin{center}
  \footnotesize{  \begin{tabular}[t]{|c||c|c|c||c|c|c|}
      \hline
      &\multicolumn{3}{|c||}{All points
      considered}&\multicolumn{3}{|c|}{last four points considered}\\\hline
      &$X$&$b$&$c$&$X$&$b$&$c$\\\hline\hline
      \multicolumn{7}{|c|}{$\beta=0.3$}\\\hline
      $\alpha$ &0.297&1.41&0.215&0.309&1.18&0.187\\\hline
      $\lambda$&0.868&0.423&0.197&0.877&0.404&0.165\\\hline
      $J/$meV   &14.3&8.88&0.142&13.9&37.4&0.269\\\hline\hline
      \multicolumn{7}{|c|}{$\beta=0.22$}\\\hline
      $\alpha$ &0.301&1.41&0.208 & 0.318&1.15&0.175\\\hline
      $\lambda$&0.867 & 0.850&0.274 & 0.886&0.526&0.175\\\hline
      $J/$meV   &13.6 &12.0&0.191 & 13.8& 15.2 &0.191\\\hline\hline
      \multicolumn{7}{|c|}{$\beta=0$}\\\hline
      $\alpha$ &0.308&1.34&0.191&0.323&1.30&0.173\\\hline
      $\lambda$&0.900&0.38&0.140&0.896&0.754&0.191\\\hline
      $J/$meV   &13.6&8.00&0.133&13.7&9.31&0.139\\\hline\hline
      \multicolumn{7}{|c|}{$\beta=-0.3$}\\\hline
      $\alpha$ &0.314&1.28&0.180&0.326&1.37&0.173\\\hline
      $\lambda$&0.913&0.369&0.127&0.903&0.832&0.192\\\hline
      $J/$meV   &13.6&7.69&0.127&13.6&9.12&0.136\\\hline
    \end{tabular}}
  \end{center}
\ TABLE II.\small{ Same as in Tab.~I based on different experimental points:\
[$(0,0); 2.1$], [$(0,0.05); 4.35$], [$(0,0.25); 15.7$], [$(1,0); 5.78$].}

\normalsize

\begin{figure}[hb]
\includegraphics[width=8.8cm]{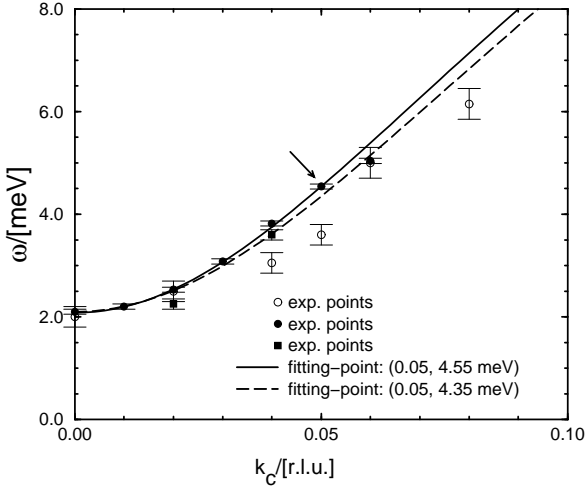}
\caption{Enlargement of Fig.~\ref{disp_big_c} for small
  wave vectors at $\beta = 0.3$. The solid curve is
  the same as in Fig.~\ref{disp_big_c}, leading to Tab.~I. The dashed
  line corresponds to a 10$^{\rm th}$ 
  order fit where the point ($k_c=0.05, \omega=4.55$meV)
  indicated by the arrow is replaced by ($k_c=0.05, \omega=4.35$meV),
  leading to Tab.~II. The filled circles correspond to highly
  accurate experimental points for which the error bars are of the
  size of the symbols.}
\label{disp_small_c}
\end{figure}

A closer inspection of Figs.~\ref{disp_big_c} and~\ref{disp_big_b}
reveals that we are confronted with a certain arbitrariness of which
fit we should favor. The experimental errors enhance this problem. The
filled circles in the range of small wave vectors 
in Fig.~\ref{disp_big_c} represent 
experimental points which have been measured with a high degree of
precision. Thus it is reasonable to fit the theoretical curve as well
as possible to these points. Fig.~\ref{disp_small_c} shows an
enlargement of this region. The solid line is the 10$^{\rm th}$
 order fit result
for $\beta=0.3$ as the solid line in
Fig.~\ref{disp_big_c}. The depicted arrow indicates the experimental
point ($k_c=0.05, \omega=4.55$meV) used to obtain Tab.~I.

A likewise well suited curve, however, is produced if one uses the
point ($k_c=0.05, \omega=4.35$meV) for the fit keeping the other
points (Tab.~II). It is not possible to prefer one of the two lines in
Fig.~\ref{disp_small_c} to the other on the basis of their agreement
to the experimental data. Hence we choose these two fits as the bounds
within which all fits are acceptable. The corresponding parameter
values $X_1$ (fit 1) and $X_2$ (fit 2) provide an interval [$X_1,X_2$]
which we expect to contain the true model parameter $\bar{X}$. 
Hence the latter is estimated by 
\[\bar{X}=(\bar{X}_1+\bar{X}_2)/2 \pm \Delta \bar{X}\]\ ,
with $\bar{X}_i=1/2(X_i^{\rm all\ points}+X_i^{\rm last\ 4\
  points})$, $\Delta \bar{X}_i = |\bar{X}_i-X_i^{\rm all\ points}|$ and
  $\Delta \bar{X} = \max \{|\bar{X}-\bar{X}_1|,\Delta \bar{X}_1,\Delta
  \bar{X}_2\}$. 
The results are summarized in Tab.~III.

\begin{center}
  \begin{tabular}[t]{|c|c||c|c|}
    \hline
    parameter&interval&parameter&interval\\\hline\hline
    \multicolumn{4}{|c|}{$\beta=0.3$}\\\hline    
    $\alpha$&0.27(4)&$\alpha_0$&0.25(3)\\\hline
    $\lambda$&0.86(2)&$\delta$&0.08(1)\\\hline 
    $\mu$&0.27(1)&$\mu_0$&0.29(1)\\\hline
    $J$/meV&13.8(5)&$J_0$/meV&12.8(6)\\\hline
    \multicolumn{4}{|c|}{$\beta=0.22$}\\\hline    
    $\alpha$&0.27(4)&$\alpha_0$&0.25(3)\\\hline
    $\lambda$&0.86(2)&$\delta$&0.08(1)\\\hline 
    $\mu$&0.21(1)&$\mu_0$&0.23(1)\\\hline
    $J$/meV&13.4(4)&$J_0$/meV&12.5(5)\\\hline
     \multicolumn{4}{|c|}{$\beta=0.0$}\\\hline    
    $\alpha$&0.27(5)&$\alpha_0$&0.25(5)\\\hline
    $\lambda$&0.88(3)&$\delta$&0.07(2)\\\hline 
    $\mu$&0.13(1)&$\mu_0$&0.14(1)\\\hline
    $J$/meV&13.2(6)&$J_0$/meV&12.4(7)\\\hline
     \multicolumn{4}{|c|}{$\beta=-0.3$}\\\hline    
    $\alpha$&0.27(5)&$\alpha_0$&0.25(5)\\\hline
    $\lambda$&0.88(3)&$\delta$&0.06(2)\\\hline 
    $\mu$&0.08(2)&$\mu_0$&0.09(1)\\\hline
    $J$/meV&13.1(5)&$J_0$/meV&12.3(7)\\\hline
\end{tabular}
\end{center}
\ TABLE III.\small{Final parameter intervals resulting from Tabs.~I and~II
  for three different values of $\beta$.}\vspace*{3mm}
\normalsize

\begin{figure}
\includegraphics[width=8cm]{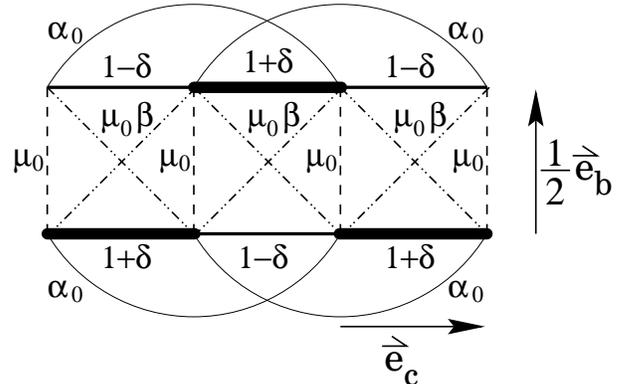}
\caption{
Alternative notation for the couplings in the dimerized phase of
CuGeO$_3$.
The couplings are denoted relative to the average nearest-neighbor coupling
$J_0$ in the chains.
\label{koppl2}}
\end{figure}

For the readers' convenience Tab.~III also gives the results in the
more commonly used parameters $\delta$, $\alpha_0$, $\mu_0$, $J_0$
and $\beta$. This notation is connected to the one used so far
in this article by
 \begin{eqnarray}\nonumber
  J=J_0(1+\delta)&,&\ \ \lambda=\frac{1-\delta}{1+\delta}\\\label{Subs_1_2}
  \alpha=\frac{\alpha_0}{1-\delta}&,&\ \ \mu=\frac{\mu_0}{1+\delta}\ .
\end{eqnarray}
It  corresponds to the Hamiltonian depicted in Fig.~\ref{koppl2}.

\section{Implications for the  Susceptibility}
The temperature dependence of the homogeneous susceptibility
$\chi(T)$ is often used to determine the parameters of CuGeO$_3$
\cite{uhrig97a,riera95,casti95,fabri98a,bouze99}. Already the 
Curie-Weiss temperature $\Theta$ provides valuable information
on the sum of the coupling constants. This is particularly useful
to detect frustration. The dispersions are governed by the difference
of the direct and the frustrating coupling whereas $\chi(T)$
at larger temperatures is more influenced by  the sum of
direct and  frustrating coupling. 

The analysis of the Curie-Weiss temperature alone bears some risks.
It is easy to calculate but difficult  to determine experimentally
since it has to be deduced from values at high temperatures where
$\chi(T)$ is fairly small and hence strongly influenced by background
 effects (van Vleck, diamagnetism) or by slight structural changes. 

\begin{figure}
\includegraphics[width=8.8cm]{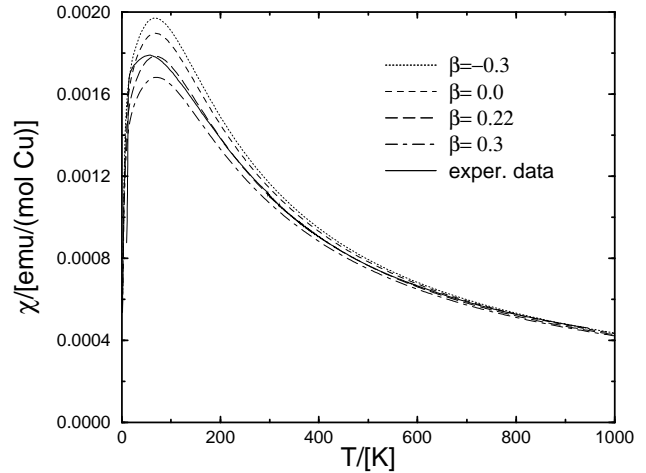}
\caption{Comparison of experimental data in b-direction
 \protect\cite{fabri98a} and theoretical 
susceptibilities for various values of the interchain frustration. The 
$g$ value used is $2.26$ \protect\cite{honda96,pilaw97}.
\label{fig:susz}}
\end{figure}

A convincing fit for temperatures above 50K is given by Fabricius
{\it et al.} in Ref.~\cite{fabri98a} on the basis of frustrated chains.
The inclusion of interchain couplings, however,
 would spoil the excellent agreement
and a re-determination of the constant would be necessary. A description
of $\chi(T)$ on the basis of a two-dimensional model has not been done
except for a consideration of the two leading coefficients in an expansion
in $1/T$ in [I]. In Fig.~\ref{fig:susz} we show the same 
high quality experimental data as in Ref.~\cite{fabri98a} and compare it
to theoretical curves at four values of $\beta$. The theoretical curves
are obtained by computing a [4,5] Dlog Pad\'e approximant $\chi_0(T)$ 
based on the
high temperature series provided in Ref.~\cite{buhle00a} 
for the frustrated chains. This procedure provides excellent
results down to $T\approx J/5$ \cite{buhle00b}.
 The asymptotic behavior of the approximant
is chosen such that $\chi_0(T)$ vanishes linearly on $T\to 0$ as is 
to be expected for a two-dimensional massless antiferromagnet.
Besides this feature the two-dimensionality is incorporated
on a chain-mean-field level
\begin{equation}
\label{eq:cmf}
\chi(T) = \frac{\chi_0(T)}{1+ 2\mu_0(1+2\beta)\chi_0(T)} \ .
\end{equation}
This relation is exact in linear order in $\mu$. Estimates of 
 corrections quadratic in $\mu$ indicate that they are negligible
for the values of $\mu$ and $\beta$ in which we are interested.

From the results in Fig.~\ref{fig:susz} it is evident that
the interchain frustration cannot be neglected. Only for a finite value
of about $0.22$ a very good agreement can be obtained. The
agreement to the frustrated chain model \cite{fabri98a} is still better
since the position of the maximum is also reproduced. But on the basis of
the neutron scattering results \cite{regna96a} it is undoubtful
that CuGeO$_3$ is a two-dimensional substance. Furthermore, it must be
considered that the previous fit \cite{fabri98a} was a two-parameter 
($J$ and $\alpha_0$) fit whereas only one parameter ($\beta$) is fitted
to obtain Fig.~\ref{fig:susz}. The other parameters ($J_0,\alpha_0,\mu_0$)
 were determined from an entirely different experiment. Hence 
 the agreement for $\chi(T)$ corroborates also the validity of 
the parameters determined in the preceding section.

\section{Discussion}

We will first discuss our results and propose a set of parameters.
 Then we will put these results into the context of other results
in the literature.

Let us consider the remaining difference between experiment and
theory concerning $\chi(T)$ in Fig.~\ref{fig:susz}.
 There are four conceivable sources
for it. The first are experimental inaccurcies. We are not in the
position to judge this aspect. We just like to comment that from
the results and error bars in Fig.~\ref{disp_small_c} it is obvious
that the experimental data is not completely consistent so that 
this explanation is possible.

Second, it is conceivable that the couplings change across the transition,
i.e.\ the intrachain frustration at $T\approx 0$ (where the dispersions are 
measured) is different from the
one above $T_{\rm SP}$ (where $\chi(T)$ is measured). 
Since we are considering a spin-Peierls 
transition there is definitely a structural change. So far, however,
the assumption that only the dimerization changes worked well. The
structural changes in the transition are very small \cite{brade96a} whereas
the changes needed to explain the discrepancy are of the order of 20 to 30 \% 
(assuming a change in the intrachain frustration).
Estimates point into the direction that the changes on the couplings
are unimportant \cite{werne99}. Yet the estimates concern in the first
place the nearest-neighbor coupling only.
Quantitative ab initio calculations of the frustration are 
very difficult \cite{rosne00}, even more so
for changes of the frustration. So, again, this explanation is perhaps
not the most promising but cannot be excluded either.

Third, the influence of the phonon dynamics is to be considered. It is shown
that spin-Peierls systems
 can be unitarily transformed in such a way that an effective spin
model remains at low energies uncoupled to the phonons \cite{uhrig98b}.
The effective couplings in a single chain model
become temperature dependent so that this may
account for the discrepancy. But it turns out for non-resonant
phonons ($\omega > J$) that these effects leave the susceptibility 
fairly unchanged. This is so since these effects become significant
 for relatively large temperatures  where $\chi(T)$ depends only on
the sum of all couplings which is unchanged by the unitary transformation
(this is observed in (VO)$_2$P$_2$O$_7$ \cite{norma00}). So this reason
appears rather unlikely even though it looked plausible at first sight.

Fourth, one has to think about any kind of 
precursors of the spin-Peierls transition. By this we mean on one side
the critical fluctuations which appear in a narrow region ($\approx 3K$)
 around the spin-Peierls temperature \cite{werne99}. On the other side,
we mean any precursor which goes beyond a purely static spin model.
Experimentally, a finite lattice correlation length can be detected
already at $T\approx 40$K far above the actual transition \cite{pouge94}.
From there on deviations from the behavior of a static spin model
should be observable. In the adiabatic limit, for instance, the fluctuation
yield already a reduction of the susceptibility \cite{dumou96}. 
What happens in the
antiadiabatic limit has not yet been investigated for a two or higher
dimensional model. The mapping in Ref.~\cite{uhrig98b} leads to
four-point interchain couplings the influence of which is unclear so far.

In view of the above mentioned possible pitfalls of the static model
 the agreement in Fig.~\ref{fig:susz} is already very convincing.
Summarizing our results we propose the parameters given for
$\beta=0.22$ in Tab.~III to be the ones deduced from the dispersion
data. Assessing the reliability of our estimates, we redo the
analysis of the susceptibility for $\alpha_0=0.28$ (the upper bound
of our estimate for $\alpha_0$) with the corresponding
value of $J_0=12.8$meV. Then the optimum $\chi(T)$ is obtained for 
$\beta=0.15$; the $\chi(T)$ curve is almost identical to the one
shown in Fig.~\ref{fig:susz}. So the value of $\beta$ cannot be determined
very precisely, but it should be in the range $\beta=0.2(1)$.
A certain dicrepancy between the optimum parameters for the
$T=0$ dispersion data and for the $\chi(T>T_{\rm SP})$ data remains.

We split the comparison of our findings to previous works into three groups.
The first comprises the analyses on the basis of a one-dimensional
model \cite{riera95,casti95,fabri98a,welle98}. The most striking difference
to the results for static spin models \cite{riera95,casti95,fabri98a}
is that the dimerization $\delta$
is not of the order of 1\% but significantly larger. This is not astounding
since it has been noted already in [I] that the gap is lowered by
the interchain hopping. Hence the neglect of the latter requires to lower
the gap otherwise, i.e.\ by assuming a lower dimerization.

Our intrachain frustration
is slightly larger than the one of Castilla {\it et al}.\ ($0.24$), but
significantly smaller than the value of Riera and Dobry ($0.36$) or the
value of Fabricius {\it et al}.\ ($0.35$). Fabricius {\it et al}.\
showed that the value $\alpha_0=0.24$ is too small for a single chain
model. The difference between the larger frustration value in the single
chain model to our value results directly from the interchain coupling.
As can be nicely seen in Eq.~(\ref{eq:cmf}) the interchain coupling lowers
the susceptibility without changing (in the chain mean-field
approximation) the position of its maximum. The one dimensional models
favor a larger intrachain frustration and a concomitant larger coupling $J$
 in order to reduce the magnitude of the susceptibility.

The claim by Wellein and coworkers that the dimerization experimentally 
found to be larger than would fit to a static 1D model \cite{buchn99a} is due
to the phonon dynamics is not compelling. They use a root-mean-square
definition of the dimerization which naturally provides larger values
for the dimerization since it includes all the fluctuations. 
The dispersion perpendicular to the chains, however, is an unambiguous
experimental fact. Furthermore, Trebst and
coworkers \cite{trebs99} do not find a substantial gap renormalization
 for parameters
relevant for CuGeO$_3$ even though one should take care of different
schemes to couple the phonons. 

 Let us turn to the second group of papers considering the essentially 
two-dimensional character of CuGeO$_3$.
The first work  used a bond-operator
technique \cite{cowle96}. No frustration was considered, hence rather small
values of $J=10.2$meV and  a rather
small interchain hopping $\mu\approx0.06$ resulted. The same technique 
was also applied later again by Brenig \cite{breni97b} including frustration.
It turned out, however, that only $\lambda(1-2\alpha)$ and $\mu(1-2\beta)$
matter on the free-boson level. Hence an independent determination of the
frustration is not possible. Using additional input ($\delta=0.012$)
the values $\alpha_0=0.059$ and $\mu(1-2\beta)=0.054$ were obtained.
 In view of 
the extensive comparisons to numerical results made in Ref.~\cite{breni97b}
it appears that the bond-operator method overestimates the influence of
additional couplings  such as dimerization or frustration. Generally,
the values for dimerization or frustration tend to be too low. This is
confirmed by our findings in the present work.

Compared to [I] ($\alpha_0=0, J=9.8$meV, $\delta=0.12, \mu_0=0.34,\beta=0.3$)
  the extended series on which our present analysis is based
gives a much better handle on intrachain frustration, see Fig.~\ref{trend_1}. 
Only
in the present work, we are able to assess its value reliably. With respect
to the interchain frustration, the present results agree qualitatively
with those in [I] where such a frustration was proposed for CuGeO$_3$ first.
 The use of  susceptibility information has been improved
in the present work since the whole $\chi(T)$ curve is used, not only
the leading coefficients.

Bouzerar {\it et al.\ } have carried out an estimate leading to results
not too far from ours: $\delta=0.065,\alpha_0=0.2, J=12.2$meV, $\mu=0.15$.
They used just 
linear order in the interchain hopping without interchain frustration
and some square-root averaging with numerical results for chains 
to describe the dispersion. The intrachain frustration ($\alpha_0=0.2$)
could only be taken from the Curie-Weiss constant. The resulting $\chi(T)$
has similarities with the experimental one.

The third group comprises ab-initio calculations of the exchange couplings
and of the spin-phonon couplings.
Microscopic calculations \cite{geert96,brade96a} 
find relatively large values of the 
dimerization between $0.07$ and $0.2$ in agreement with our findings.
(Even though there is also a different result \cite{feldk99})
 Very important for our work  are recent
results by Drechsler and coworkers supporting the existence of 
interchain frustration \cite{rosne00}. Werner and coworkers \cite{werne99}
estimate a large dimerization from the spin-phonon couplings and the
 shifts of the ions ($\delta=0.11$).
From the balance of elastic and magnetic energy in the D phase
they obtain without frustration a lower bound of
$\delta>0.044$. Assuming critical frustration $\alpha_0=0.2412$ 
they find even
$\delta>0.078$ which fits very nicely to our findings.

In summary, we provide by the present work a determination in great detail of
the coupling parameters ($\beta=0.2(1)$ and right column in 
Tab.~III under $\beta=0.22$)
 of CuGeO$_3$ based on a static dimerized
spin model at $T=0$. The experimental input comes from inelastic
neutron scattering. The implications of the parameters found for the
susceptibility are also studied. Very good agreement could be obtained
fitting the interchain frustration appropriately. A small discrepancy at
low temperatures around $50$K
indicates that the static spin model is probably insufficient
to describe CuGeO$_3$ completely.
By this work, we proved the outstanding possibilities of high-order
series expansions (around the dimer limit or around the limit $T=\infty$)
in the analysis of experimental data.

\section*{Acknowledgements}
The authors like to thank B.~B\"uchner, A.~B\"uhler,
U.~L\"ow, B.~Mari\'c,
E.~M\"uller-Hartmann, and F.~Sch\"onfeld for fruitful discussions. 
The provision of the experimental
data by B.~B\"uchner, T.~Lorenz and by 
L.~P.~Regnault is gratefully acknowledged. Furthermore, we thank
the Regional Computing Center of the University of Cologne for its
kind and efficient support. This work was
supported by the Deutsche Forschungsgemeinschaft in the 
Schwerpunkt 1073 and in the Sonderforschungsbereich 341.

\begin{thebibliography}{10}
\bibitem{note-corr}
Some mixed third order 
hopping amplitudes in Ref.~\cite{uhrig97a}, Eq.~(2)) are
incorrect: \newline
$t_0$:
$+(8\overline\alpha-16\overline\beta+18\overline\alpha\overline\beta
+9\overline\beta^2-12\overline\alpha\overline\beta^2)\lambda\mu^2/32$ 
\newline should be
$+(8\overline\alpha-16\overline\beta
+15\overline\alpha\overline\beta^2)\lambda\mu^2/32$ \newline
$+\lambda^4/128$ should be $-\lambda^4/128$ \newline
$-3\lambda^2\mu^2/16$ should be $-\lambda^2\mu^2/32$. \newline
$t_{10}$: $+2\overline\alpha\overline\beta^2\lambda\mu^2/32$ 
should be $+\overline\alpha\overline\beta^2\lambda\mu^2/64$ \newline
$+\lambda^2\mu^2/32$ should be $+3\lambda^2\mu^2/64$. \newline
$t_{11}$: $+7\overline\alpha^2\overline\beta\lambda^2\mu/128$ should
be $+4\overline\alpha^2\overline\beta\lambda^2\mu/128$ \newline
$+7\overline\beta^3\mu^3/128$ should be
$+4\overline\beta^3\mu^3/128$\newline
$+5\lambda^2\mu^2/128$ should be $+9\lambda^2\mu^2/128$. \newline
$t_{21}$: $-5\overline\alpha^2\overline\beta/\lambda^2\mu/256$ should
be $-2\overline\alpha^2\overline\beta\lambda^2\mu/256$ \newline
$-3\overline\beta^3\mu^3/256$ should be $-3\overline\beta^3\mu^3/128$.
\end{thebibliography}

\end{document}